\documentclass[twocolumn,showpacs,secnumarabic,amssymb,nobibnotes,aps,prl]{revtex4-1}
\usepackage{amssymb}
\usepackage{graphicx}
\usepackage{color}

\usepackage{url,hyperref}
\usepackage[usenames,dvipsnames]{xcolor}
\hypersetup{colorlinks=true, linkcolor=BrickRed, urlcolor=blue!50!black, citecolor=blue!50!black}
\usepackage[capitalize]{cleveref}
\usepackage{cleveref}
\crefrangelabelformat{equation}{(#3#1#4)$-$(#5#2#6)}

\begin{document}
\title{Active Particles in Explicit Solvent: Dynamics of clustering for alignment interaction}

\author{Arabinda Bera$^{1,2,{\dagger}}$, Soudamini Sahoo$^{3,{\dagger}}$, Snigdha Thakur$^{3,*}$, and Subir K. Das$^{1,2,*}$}

\affiliation{$^1$Theoretical Sciences Unit, Jawaharlal Nehru Centre for Advanced Scientific Research, Jakkur P.O, Bangalore 560064, India}

\affiliation{$^2$School of Advanced Materials, Jawaharlal Nehru Centre for Advanced Scientific Research,
Jakkur P.O, Bangalore 560064, India}
 
\affiliation{$^3$Department of Physics, Indian Institute of Science Education and Research Bhopal, Madhya Pradesh 462066, India}

\date{\today}

\begin{abstract}
We study dynamics of clustering in systems containing active particles that are immersed in an explicit solvent. For this purpose we have adopted a hybrid simulation method, consisting of molecular dynamics and multi-particle collision dynamics. In our model, overlap-avoiding passive interaction of an active particle with another active particle or a solvent particle has been taken care of via variants of Lennard-Jones potential. Dynamic interaction among the active particles has been incorporated via the Vicsek-like self-propulsion that facilitates clustering. We quantify the effects of activity and importance of hydrodynamics on the dynamics of clustering via variations of relevant system parameters. We work with low overall density of active particles. For this the morphology consists of disconnected clusters, the mechanism of growth switching among particle diffusion, diffusive coalescence and ballistic aggregation, depending upon the presence or absence of active and hydrodynamic interactions. Corresponding growth laws have been quantified and discussed in the background of appropriate theoretical pictures. Our results suggest that multi-particle collision dynamics is an effective method for investigation of hydrodynamic phenomena even in active matter systems.
\end{abstract}
\maketitle
\section{Introduction}
Be it a colony of bacteria or a herd of sheep, clustering in active matter systems, containing self-propelling particles \cite{cates2015,link2012,palacci2013,kumar2014,vicsek1995,marchetti2013,ramaswamy2015}, is rather common. Phenomena associated with such assemblies received much attention in the passive scenario \cite{binder1991,bray2002,onuki2002,puri2009,hohenberg1977,lifshitz1961,fisher1967,binder1974,binder1977,tanaka1996,siggia1979,roy2019,roy_sm2013,roy2012,roy_jcp2013,das2012}. Currently, for experimentalists and theorists alike, the focus is on the active counterpart \cite{cates2015,link2012,palacci2013,kumar2014,vicsek1995,marchetti2013,ramaswamy2015,wysocki2014,belmonte2008,caprini2020,bechinger2016,das2017,das2014,trefz2016,siebert2018,thakur2012,yu2018,thakur2011}. In the theoretical literature, particularly in the domain of statistical mechanics \cite{palacci2013,kumar2014,marchetti2013,ramaswamy2015}, there is strong interest in the construction of meaningful methods, to help understand these clustering phenomena. Like in the passive systems, for active matters also, associated interest is rather general. This includes topics of first and second order phase transitions \cite {wysocki2014,belmonte2008,caprini2020,bechinger2016,das2017,das2014,trefz2016,siebert2018,thakur2012,yu2018,thakur2011}. Steady states in active matter systems are counterparts of equilibrium situations in passive systems. The broad objectives, in such a limit or on approach to it, are to identify universality in critical and coarsening phenomena \cite{bray2002,hohenberg1977,fisher1967,das2012}.

In passive systems it is well established that hydrodynamics plays important roles in both equilibrium \cite{hohenberg1977} and nonequilibrium \cite{bray2002,binder1974,siggia1979,roy2012,das2012} dynamics. E.g., in presence of hydrodynamics, progress of phase separation, following a quench of a homogeneous system inside the coexistence region, occurs much faster. Such effects may be expected in the active case also. Understanding of these requires effective simulation techniques that can take well care of hydrodynamic conservations in computers, while dealing with large numbers of active particles and adequately high density of solvent. In this letter we consider a collection of Vicsek-like \cite{vicsek1995} active particles immersed in an explicit solvent. The overall dynamics is carried out via an advanced combination of multi-particle collision dynamics (MPCD) \cite{kapral2008} and molecular dynamics (MD) \cite {frenkel2002,allen1987} techniques. To the best of our knowledge, phase separation models having explicit solvent are very limited even for passive systems~\cite{furukawa2010,tanaka2000,sdas2018,winkler2013,li2015}, leave aside the implementation of an advanced hydrodynamic method for studying large scale phenomena exhibited by active microswimmers~\cite{navarro2015,navarro2014,li2015,zottl2014,blaschke2016}. Our study fills this important gap, in addition to obtaining important simulation results and explaining these via an appropriate theoretical picture.

In passive fluids, the influence of hydrodynamics on the evolution or growth rate, during phase separation, depends on the overall density or composition of particles \cite {bray2002,binder1974,tanaka1996,siggia1979,roy2019,roy_sm2013,roy2012,roy_jcp2013,das2012}. For low overall density, say, in a vapor-liquid transition, the evolving morphology consists of clusters (of particles) that are disconnected from each other \cite{roy_sm2013,roy2012,roy_jcp2013}. Growth of such a morphology is expected to occur via coalescence of the clusters \cite {binder1974,binder1977,siggia1979,roy2012,midya2016}. Here, we investigate structure and dynamics in such a region of the phase diagram, for active as well as passive systems, the emphasis being on the former. We observe that the method leads to appropriate hydrodynamic mechanisms. In the passive case this provides growth via diffusive coalescence (DC) mechanism which is in good quantitative agreement with a picture proposed by Binder and Stauffer \cite {binder1974}. When the activity is turned on we show that the growth occurs via the ballistic aggregation (BA) mechanism \cite {midya2016,carnevale1990,trizac2003,trizac1996,brilliantov2015,paul2017,paul2018}. The emerging growth law has been quantified and explained via an appropriate theoretical consideration \cite {carnevale1990}.

During phase separation average mass, $M$, of domains or clusters typically evolves \cite {bray2002} with time ($t$) as  $M \sim t^{\beta}$. For disconnected morphology, in presence of hydrodynamics, one expects $\beta=1$ for DC of clusters \cite {binder1974,tanaka1996,siggia1979,roy2012}. The expectation is different for BA \cite {carnevale1990,trizac2003,trizac1996,brilliantov2015,paul2017,paul2018}. For the latter, the value of ${\beta}$ should be  $1.2$ in space dimension $d=3$ \cite{carnevale1990,trizac2003,trizac1996,brilliantov2015,paul2017,paul2018} that may, however, increase when the velocities of the clusters are correlated \cite {midya2016,carnevale1990}. For each of these mechanisms, $\beta$ can take different values when the structures are non-spherical \cite {midya2016}.
\section{Model and Methods}
In our model $N_a$  active particles, each of mass $m_a$ and diameter ${\sigma}$, are immersed in a solvent consisting of $N_s$ point-like particles, each having mass $m_s$. The overlap-preventing interaction between two active particles, a distance $r$ apart, is implemented via a shifted and force-corrected Lennard-Jones potential~\cite{allen1987}, having a cut-off radius $r_c=2.5{\sigma}$ and interaction strength $\epsilon$. The self-propulsion is achieved via the Vicsek interaction \cite{vicsek1995,das2017}, i.e., an active particle $i$ experiences a dynamic force ${{\bf{f}}_i}=f_a{\hat{n}}$, due to its neighbors, implemented in such a way that only its direction of motion is changed. Here, $\hat{n}$  ($=\sum_j{\bf{v}}_j/|\sum_j{\bf{v}}_j|$, ${\bf{v}}_j$ being the velocity of particle $j$ that falls within the neighborhood defined by the distance $r_c$ from particle $i$) is the mean direction of motion of neighbors and $f_a$ is the strength of active force. The interaction between the active and solvent particles, for $r<2^{1/6}({\sigma}/2)$, is governed by the Weeks-Chandler-Andersen potential \cite{weeks1971}: $U_{s}(r)=4\epsilon[(\sigma/2r)^{12}-(\sigma/2r)^6+1/4]$, the factor $1/2$ modifies the interaction diameter to $\sigma/2$ in this case. In our mesoscopic description, the interaction among the solvent particles are incorporated by the MPCD technique~\cite {kapral2008,gompper2008}, while the rest are through MD.

In the hybrid MD-MPCD simulations, the positions and velocities of all the particles are updated via the velocity Verlet algorithm \cite{frenkel2002} with a time step ${\Delta}t$. The multi-particle collisions for solvent are performed periodically via momentum exchange among particles within a cubic cell of side $\sigma$, at time intervals $\tau$. This is achieved via rotations of the relative velocity ${\delta}\bf{v}$ of $k^{th}$ solvent particle $({\bf{v}}_{s,k})$ with respect to the velocity of the centre of mass of the corresponding cell $({\bf{v}}_{s,{cm}})$, about a randomly selected axis, by an angle $\gamma$: ${\bf{v}}_{s,k}={\bf{v}}_{s,{cm}}+{\bf{R}}(\gamma){\delta}{\bf{v}}$. Here ${\bf{R}}$ is a rotation matrix, application of which keeps the mass, momentum and energy conserved. A random grid shifting is implemented in every collision step to restore the Galilean invariance  in a small mean free path limit ($\lambda<\sigma$) \cite {kapral2008,gompper2008,ihle2001}. The simulations are performed in a periodic cubic box of side $48\sigma$. We choose $m_a=8m_s$, set by considering certain solvent density equivalence. The quenched temperature of the system is $T=0.6{\epsilon}/k_B$, $k_B$ being the Boltzmann constant. This falls well inside the miscibility gap. We set $\rho_s$, the solvent number density, at $10/{\sigma}^3$ and $\rho_a$, the active particle density, at $0.05/{\sigma}^3$. Furthermore, we have ${\Delta}t=0.005t_0$ $(t_0=\sqrt{{\sigma}^2m_s/\epsilon}~)$, $\tau=0.05t_0$ and $\gamma=130^{\circ}$. The above chosen \cite{padding2006} values of various parameters ensure the fluidic behavior of solvent, with Reynolds number $Re<1$ and Schmidt number $Sc\sim10^2$. For convenience, we fix $m_s$, $\sigma$, $\epsilon$ and $k_B$ to unity. All quantitative results correspond to averaging over more than $80$ independent initial realizations.

\begin{figure}[h]
\includegraphics[width=\linewidth]{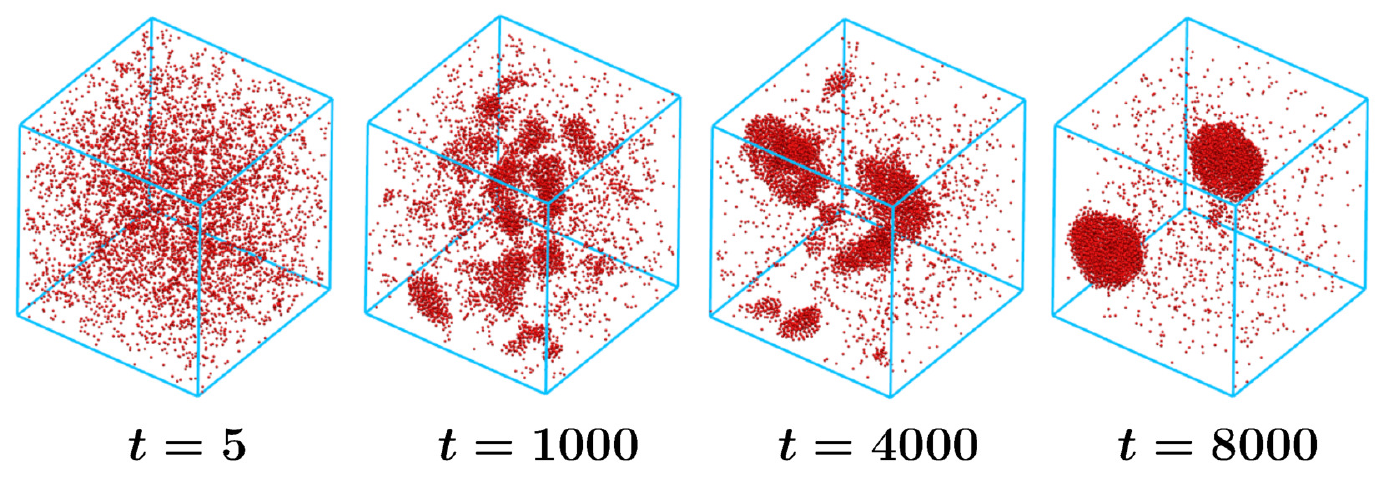}
\caption{Snapshots, recorded during the evolution of the model system, having active particles in hydrodynamic environment, are presented from different times. Locations of only the active particles are marked. These and other results for nonzero activity correspond to $f_a=3$.}
\label{snap_shot_of_colloids}
\end{figure}

\begin{figure}[h]
\centering
\includegraphics[width=\linewidth]{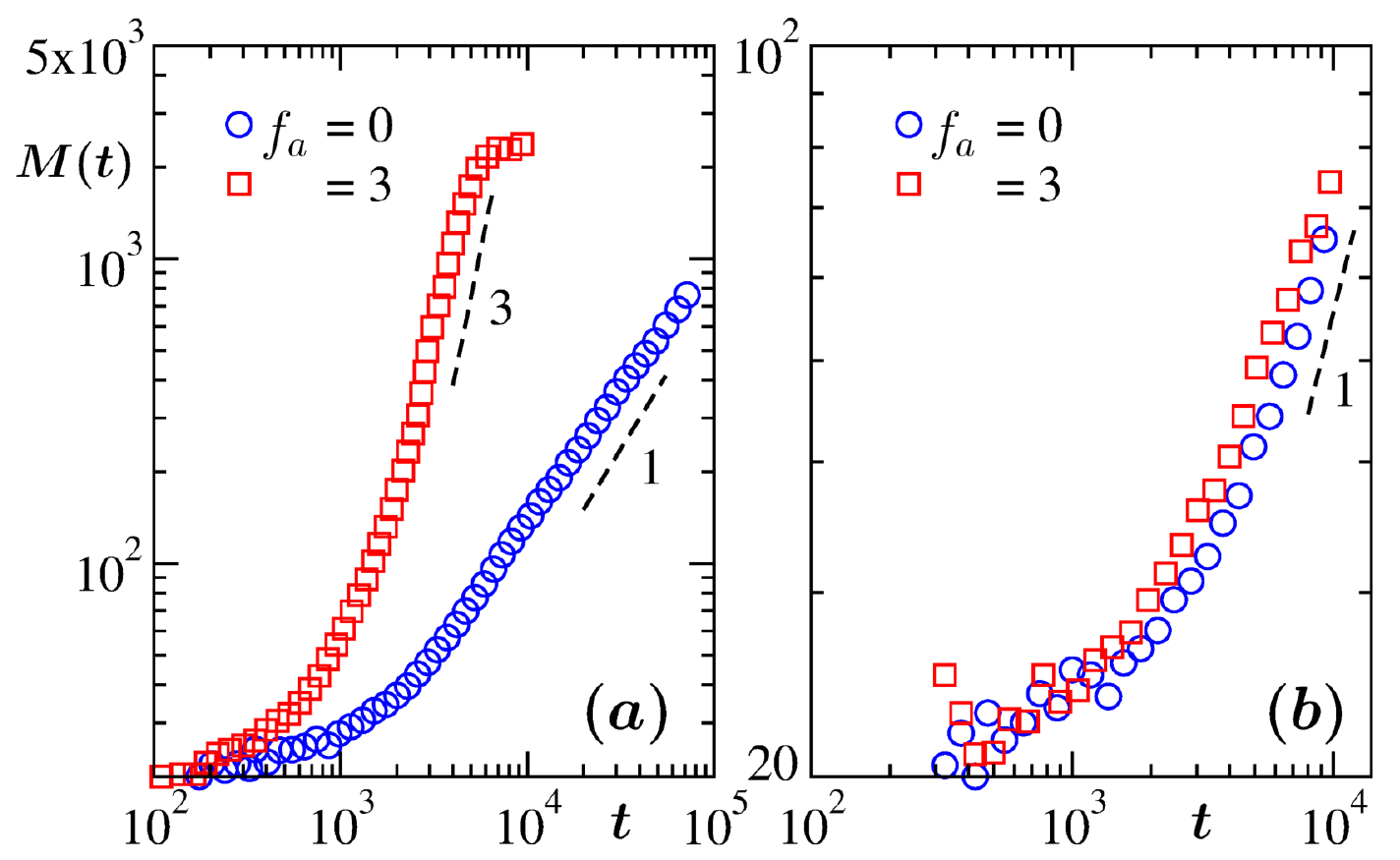}
\caption{Plots of the average mass, $M$, of the clusters versus time $(t)$, are shown on a log-log scale. We have included data that were obtained in presence of hydrodynamics (a) and in absence of hydrodynamics (b). Results for both passive ($f_a=0)$ and active ($f_a=3$) cases are shown. The dashed lines represent power-laws with mentioned values of the exponents.}
\label{mass_growth}
\end{figure}

\begin{figure}[h]
     \centering
      \includegraphics[width=\linewidth]{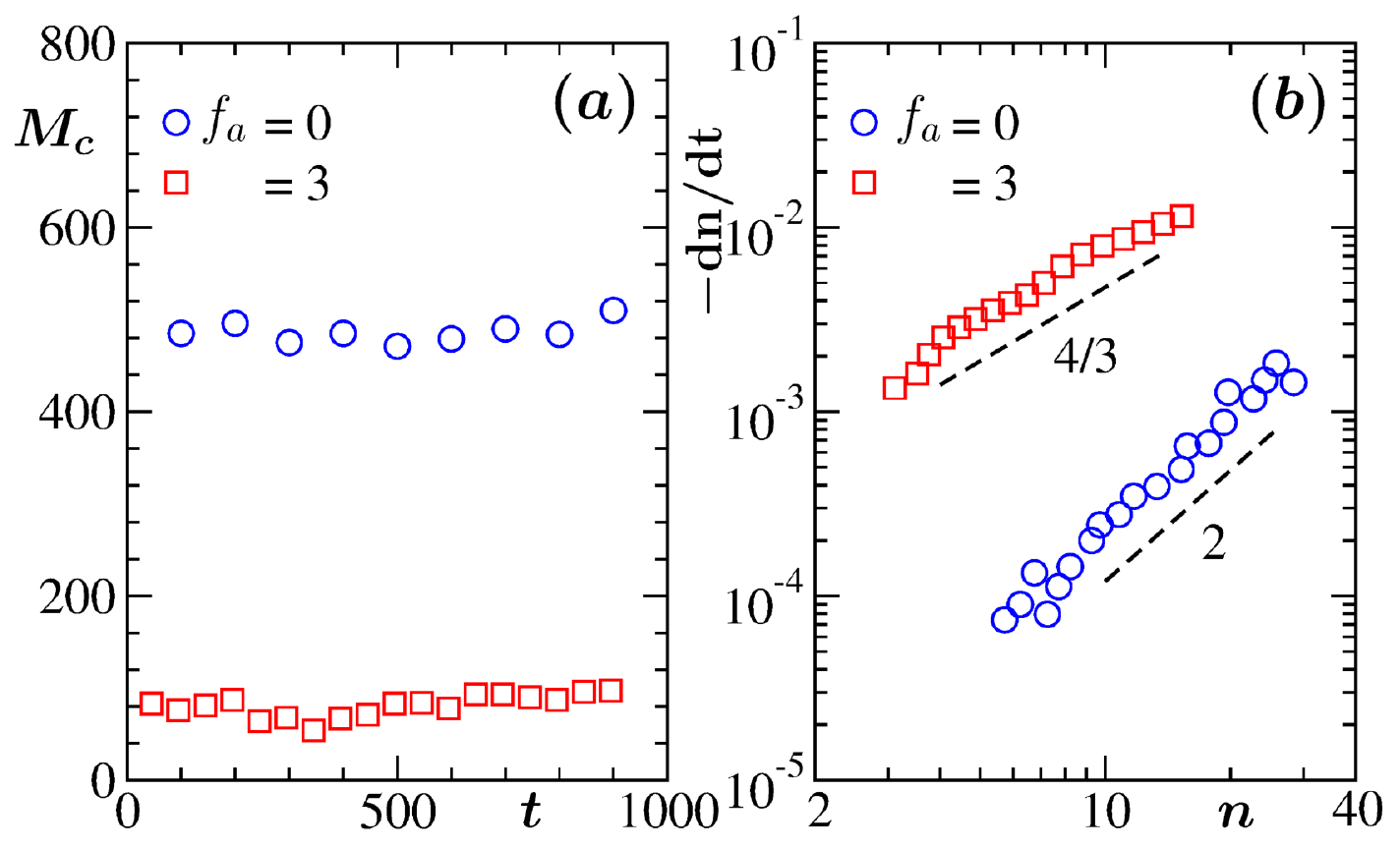}
     \caption{(a) Plots of masses in two clusters, one each from $f_a=0$ and $f_a=3$, are shown as a function of time. These were obtained by following the cluster trajectories for extended periods of time. (b) Log-log plots of $-\rm{d}n/\rm{d}t$ versus $n$, for $f_a=0$ and $3$. The dashed lines represent power-laws, exponents for which are mentioned next to them. These results correspond to the situation when hydrodynamics was on.} 
     \label{nop_cluster}
\end{figure}

\begin{figure}[h]
     \centering
   \includegraphics[width=\linewidth]{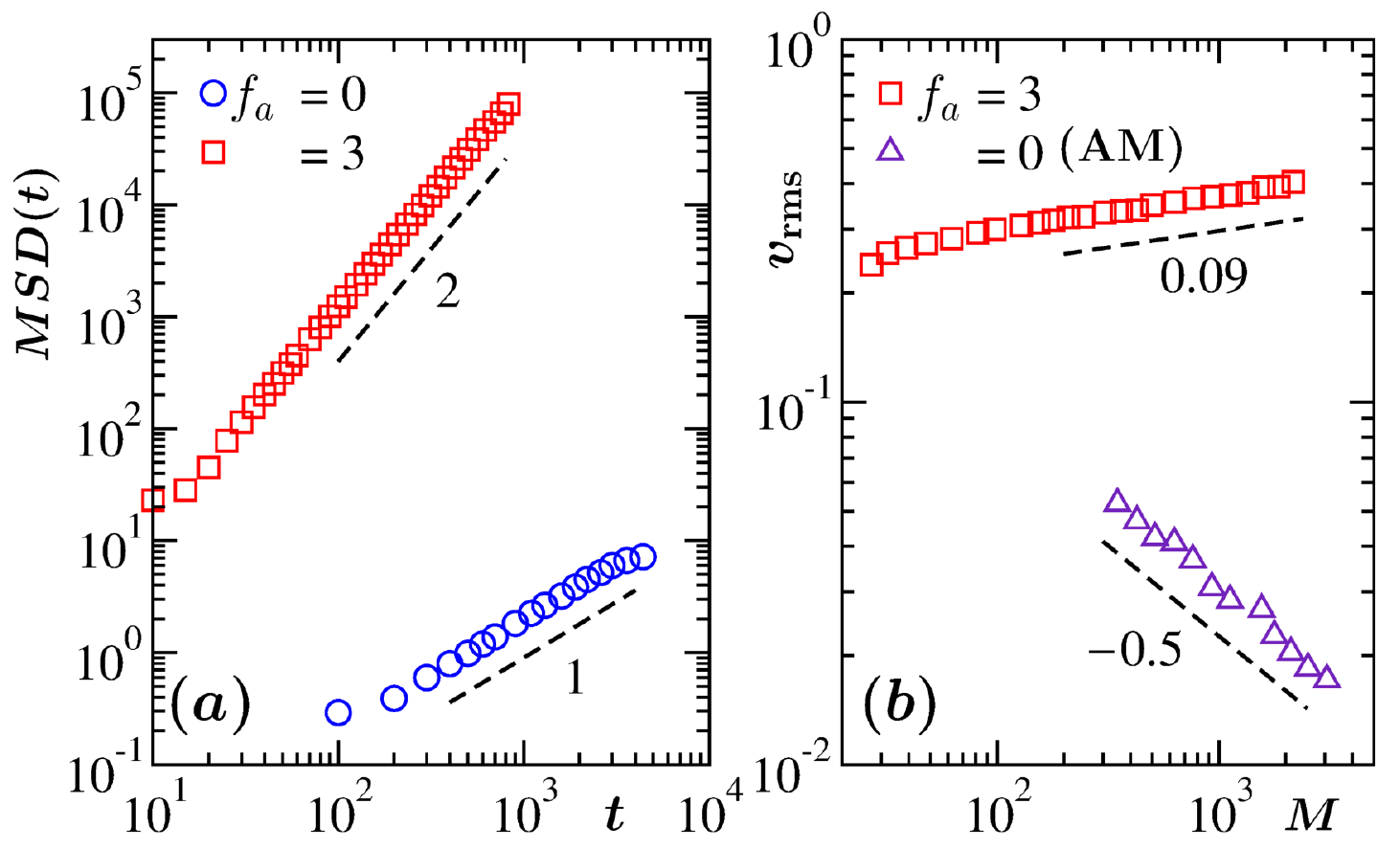}   
     \caption {(a) Plots of mean-squared-displacement (of clusters) versus time, on a log-log scale. Typical results for passive and active cases have been included. The dashed lines represent power-laws. Corresponding exponents have been mentioned in appropriate places. (b) Root-mean-squared velocity of clusters is plotted versus average mass of clusters, on a double-log scale. For comparative purpose data from another model (AM), showing $z=-1/2$, expected for uncorrelated motion of particles, are also included. See Ref. \cite{midya2016} for the details on this model. These results correspond to hydrodynamic environment.}
     \label{rms_cluster}
\end{figure}

\section{Results}
In Fig. \ref{snap_shot_of_colloids} we show snapshots that are obtained during the evolution following temperature quench of a system containing active particles in hydrodynamic environment. Formation and growth of clusters can be easily identified. Our objective is to quantify the growth and obtain a picture by comparing it with the corresponding outcome for its passive counterpart. To elucidate the role of hydrodynamics further, we will also present results by turning the local momentum conservation constraint off. 

In Fig. \ref{mass_growth} we show the growth plots for two values of $f_a$, viz., $0$ and $3$, $f_a=0$ being the passive case. Results in part (a) were obtained in the presence of hydrodynamics. On the other hand, growth pictures, in absence of hydrodynamics, are presented in part (b). In both (a) and (b), the growths for the active case occur faster. In presence of hydrodynamics, the effects of activity, however, is striking. We believe that when hydrodynamics is turned off, via  random exchange of momenta without following the local momentum conservation requirement \cite{ali2004,li2015}, the growth, for both $f_a=0$ and $3$, should occur via the Lifshitz-Slyozov \cite{lifshitz1961} ``simple" particle diffusion mechanism, for which $\beta=1$ in $d=3$. The simulation data are consistent with this picture. In the rest of the paper we focus only on the hydrodynamic case. Note that the average mass from the snapshots were calculated in standard fashion as the first moment of the mass distribution function. Mass of an individual cluster, $M_c$, was obtained, in units of $m_a$, by appropriately identifying the cluster boundary and calculating the number of particles inside it \cite{roy2019,roy_sm2013,roy2012,roy_jcp2013}.

The hydrodynamic mechanism in the passive case provides a linear growth, i.e., $\beta \simeq 1$. This is consistent with the discussion above on DC. This value of $\beta$ is the outcome of the solution of the dynamical equation \cite{binder1974,binder1977,tanaka1996,siggia1979,roy2019,roy_sm2013,roy2012,roy_jcp2013} ${\rm{d}}n/{\rm{d}}t=-Cn^2$, for the droplet density $n$. Here $C$ is a constant, arising from the generalized Stokes-Einstein-Sutherland relation \cite{das2007}. This particular solution is obtained after imposition of the mass conservation constraint $M \propto 1/n$. On the other hand, for the active case, as suggested by the plot in Fig. \ref{mass_growth}(a), we have $\beta \simeq 3$. The reason for this value cannot be straight-forwardly guessed. Below we proceed systematically to arrive at correct understanding for both the cases. In Fig. \ref{nop_cluster}(a) we show plots for masses in two typical clusters \cite{roy_sm2013,roy2012,roy_jcp2013}, one each from $f_a=0$ and $3$, versus time. The times have been translated with respect to the beginnings of the  observations and periods are chosen in such a way that within these intervals the considered clusters did not collide with any other cluster. Near-constant values in both the cases, over long periods, imply that the growths are not happening via the simple particle diffusion mechanism. Hence, it is the coalescence of droplets \cite{binder1974,midya2016,carnevale1990,trizac2003,trizac1996,brilliantov2015,brilliantov2006,roy2018}, via one or the other mechanism, that is responsible for growth in each of the cases. 

In Fig.~\ref{nop_cluster}(b) we show plots of $-{\rm{d}}n/{\rm{d}}t$ versus $n$, on a double-log scale, for both the values of $f_a$, by restricting ourselves within the post-nucleation regimes. It appears that, for the passive case, the quadratic trend of $-{\rm{d}}n/{\rm{d}}t$ in the simulation data hints towards the presence of DC mechanism. The data set for the active case exhibits a behavior that is significantly weaker than quadratic.

In Fig.~\ref{rms_cluster}(a) we show mean-squared-displacement (MSD) of typical clusters, with the variation of time \cite{hansen2008}. Again, results are included from both active and passive cases, and the times are shifted, like in Fig. \ref{nop_cluster}(a). For the passive case the time dependence is linear. This confirms that the growth is indeed occurring via the DC mechanism \cite{binder1974,binder1977,tanaka1996,siggia1979,roy2019,roy_sm2013,roy2012,roy_jcp2013,hansen2008}. On the other hand, the quadratic enhancement in the active case is suggestive of the BA mechanism \cite{midya2016,carnevale1990,trizac2003,trizac1996,brilliantov2015,paul2017,paul2018}. For BA, the simple dynamical equation for binary collisions reads  ${\rm{d}}n/{\rm{d}}t=-{\sigma}_{\text{coll}} v_{\text{rms}}n^2$, where ${\sigma}_{\text{coll}}$ is the collision cross-section and $v_{\text{rms}}$ is the root-mean-squared velocity of the clusters \cite{carnevale1990, hansen2008}. Typically, $v_{\text{rms}}$ has power-law dependence on mass as \cite{carnevale1990} $v_{\text{rms}}\sim M^z$. By taking ${\sigma}_{\text{coll}}\sim{M}^{(d-1)/d}$, one has $\beta=d/(1-dz)$. For $\beta \simeq 3$, one expects $z \simeq 0$. This is a value much different from $-1/2$ that is observed for uncorrelated motion of clusters \cite{midya2016}. In Fig.~\ref{rms_cluster}(b) we have presented $v_{\text{rms}}$ versus $M$ plots. There we have included data for $f_a=3$ for the present hydrodynamic model as well as from another passive LJ system \cite{midya2016}. In the latter case, $z \simeq -1/2$, for $T=0.3$ and $\rho=0.05$. On the other hand, in our active case, the absolute value of $z$ is very small, though not zero. The deviation that is observed, from the expectation, can be due to the fact that the actual growth is slightly different from $t^3$. Furthermore, errors in the estimation of various other exponents will contribute to the final theoretical outcome. Nevertheless, the overall picture is quite consistent and supports the observation in Fig. \ref{nop_cluster}(b). Essentially, the set of information captured in Figs.~\ref{rms_cluster}(a) and (b) are in good quantitative agreement with the power law exponent $4/3$ in $-{\rm{d}}n/{\rm{d}}t$. Hence, it supports the simple theory of BA mechanism, for the active system.

\section{Conclusion}
We have studied kinetics of clustering in systems of Vicsek-like active particles in hydrodynamic environment. We use a combination of molecular dynamics and multi-particle collision dynamics methods for this purpose. As is well known, the Vicsek activity favors phase separation. For the  chosen final temperature and the density of the active particles, the non-equilibrium morphology consists of disconnected clusters \cite {roy2012}. This choice has the advantage of checking the effects of hydrodynamics in an efficient manner. If the clusters are immersed in a high density solvent it is expected that between collisions these will exhibit diffusive motion, at least in the passive case. Even though the outcome of such diffusive coalescence has been demonstrated in computer simulations \cite {roy_sm2013,roy2012,roy_jcp2013}, such studies with explicit solvent, like the one considered here, have never been performed earlier. 

In addition, here we asked a more advanced question: How does the motion of clusters get changed when the constituents are active? It appears that for active constituents of Vicsek type the clusters eventually move ballistically. We have quantified the growth law for this and explained it by obtaining accurate knowledge on the mass dependence of the velocities of the clusters \cite {midya2016,carnevale1990}. By tuning the system parameters \cite {ali2004} we have checked that our model provides appropriate hydrodynamic environment for both passive and active cases. For both the situations such large scale study of phase separation with explicit solvent is first in the literature, to the best of our knowledge, for the considered proximity to the coexistence curve. This work provides a quantitative knowledge of the growth phenomena observed in active systems subjected to real hydrodynamic environment and can be of much industrial significance like designing advanced materials and devices.

\section{Acknowledgement}
SKD and AB acknowledge useful discussions with R. Kapral and are thankful to the Department of Biotechnology, India, and Science and Engineering Research Board (SERB) of Department of Science and Technology (DST), India, for partial financial supports via Grant Nos. LSRET-JNC/SKD/4539 and MTR/2019/001585, respectively. ST and SS acknowledge a joint collaborative visit to JNCASR, and AB to IISER Bhopal. The computational works were performed at the HPC facilities in IISER Bhopal and JNCASR. ST acknowledges SERB, DST for funding via Grant No. SERB/PHY/EMR/2017/000922.  

$\dagger$: Equal contributors; $*$: Emails of corresponding authors -- sthakur@iiserb.ac.in, das@jncasr.ac.in

\end{document}